\documentclass[aps,prb,reprint,superscriptaddress,longbibliography]{revtex4-1}
\usepackage{amssymb}
\usepackage[utf8]{inputenc}
\usepackage{graphicx}
\usepackage{bm,color,subfigure,amsmath}

\begin{document}

\title{Hydrodynamic description of long-distance spin transport through noncollinear magnetization states: the role of dispersion, nonlinearity, and damping}

\author{Ezio~Iacocca}
\email{ezio.iacocca@colorado.edu}
\affiliation{Department of Applied Mathematics, University of Colorado, Boulder, Colorado 80309, USA}

\author{Mark~A.~Hoefer}
\affiliation{Department of Applied Mathematics, University of Colorado, Boulder, Colorado 80309, USA}

\begin{abstract}
  Nonlocal compensation of magnetic damping by spin injection has been theoretically shown to establish dynamic, noncollinear magnetization states that carry spin currents over micrometer distances. Such states can be generically referred to as dissipative exchange flows (DEFs) because spatially diffusing spin currents are established by the mutual exchange torque exerted by neighboring spins.  Analytical studies to date have been limited to the weak spin injection assumption whereby the equation of motion for the magnetization is mapped to hydrodynamic equations describing spin flow and then linearized. Here, we analytically and numerically study easy-plane ferromagnetic channels subject to spin injection of arbitrary strength at one extremum under a unified hydrodynamic framework. We find that DEFs generally exhibit a nonlinear profile along the channel accompanied by a nonlinear frequency tuneability. At large injection strengths, we fully characterize a novel magnetization state we call a contact-soliton DEF (CS-DEF) composed of a stationary soliton at the injection site, which smoothly transitions into a DEF and exhibits a negative frequency tuneability. The transition between a DEF and a CS-DEF occurs at the maximum precessional frequency and coincides with the Landau criterion: a subsonic to supersonic flow transition.  Leveraging the hydraulic-electrical analogy, the current-voltage characteristics of a nonlinear DEF circuit are presented.  Micromagnetic simulations of nanowires that include magnetocrystalline anisotropy and non-local dipole fields are in qualitative agreement with the analytical results. The magnetization states found here along with their characteristic profile and spectral features provide quantitative guidelines to pursue an experimental demonstration of DEFs in ferromagnetic materials and establishes a unified description for long-distance spin transport.
\end{abstract}

\maketitle

\section{Introduction}

Noncollinear magnetization states represent a new paradigm for the
transport of spin currents over micrometer
distances~\cite{Sonin2010,Takei2014,Iacocca2017,Iacocca2017b,Iacocca2017d,Hill2018,Takei2014b,Sonin2018,Stepanov2018,Schneider2018}. A
key concept that has enabled the study of these states is the
hydrodynamic interpretation of magnetization dynamics, originally
proposed in the seminal paper by Halperin and
Hohenberg~\cite{Halperin1969} in the context of the spin wave
dispersion relation for ferromagnets and antiferromagnets. Almost four
decades later, a similar fluid-like interpretation was used to
identify the relationship between an infinite-length, static
noncollinear magnetization state in easy-plane ferromagnets and
dissipationless spin transport~\cite{Konig2001}. These states were
characterized by a homogeneous normal-to-plane magnetization and a
winding in-plane magnetization. More importantly, energy dissipation
via damping was inoperative because the texture was assumed to be
static. As a consequence, the mutual exchange torque exerted by
neighboring spins could be interpreted as an equilibrium spin current
or exchange flow~\cite{Bruno2005} that did not exhibit any
dissipation.

While the prospect of a dissipationless spin current is tantalizing
for novel energy-efficient
applications~\cite{Chen2014,Kim2015,Kim2016b,Liu2016,Hill2018,Sprenger2018},
any magnetization dynamics are subject to dissipation via magnetic
damping~\cite{Gilbert2004}. An example is the interface between a
magnetic material and a spin sink that results in spin
pumping~\cite{Tserkovnyak2002b}. To circumvent this problem, it is
necessary to introduce energy into the system. From an analysis of the
linearized hydrodynamic equations for a ferromagnet, it was predicted
that spin injection at one extremum of a one-dimensional channel could
sustain a dynamic, noncollinear magnetization state that was termed a
\emph{spin superfluid}~\cite{Sonin2010,Takei2014}.  Despite the fact
that this is a solution to the linearized, long-wavelength
hydrodynamic equations, the magnetization vector itself exhibits fully
nonlinear spatio-temporal excursions in the form of complete planar
rotations.  As we will later show, this solution results from a
linearized analysis of the equations of motion.  The usage of the term
superfluid was borrowed from a similarity between the order parameters
that describe spin transport in a magnet and mass transport in, e.g.,
superfluid He$^4$ as well as the fact that the normal-to-plane
magnetization is approximately constant along the channel, although
very small. However, this so-called spin superfluid experiences energy
loss via a spatially diffusing spin current, yet its uniform
precessional frequency and linearly decaying spin current profile
present potential advantages to the exponential decay property of
magnons. Similar states have been predicted for
antiferromagnets~\cite{Takei2014b,Qaiumzadeh2017,Sonin2018,Ochoa2018}
and their experimental evidence in such materials has been recently
presented~\cite{Stepanov2018,Yuan2018}.

In order to avoid potential misinterpretation of the term spin
superfluid and to emphasize the nonlocal compensation of damping along
the channel by the exchange torque that originates from spin injection
at the device boundary, we will refer to spin superfluids and their
generalizations as \emph{dissipative exchange flows or DEFs} for
short.

A more realistic setting for easy-plane ferromagnetic materials must
consider the effect of in-plane anisotropy that breaks axial
symmetry. For this configuration, it was shown that the hydrodynamic
equations of motion map to a damped sine-Gordon equation, with a
nonlinear term proportional to the in-plane anisotropy
strength~\cite{Sonin2010,Iacocca2017d}. Because of the broken symmetry imposed by in-plane anisotropy, the
structure of a DEF is that of a translating train of N\'{e}el domain
walls or a soliton lattice with the same chirality and whose
inter-wall spacing increases as each domain wall propagates from the
spin injection edge to the opposite free spin edge. In the limit of
vanishing anisotropy, the train of domain walls smooths into a
sinusoidal profile, equivalent to the previously studied, axially
symmetric case~\cite{Sonin2010,Takei2014}.

The most striking feature of a DEF is that its spatial structure and
coherent precessional frequency depend on the length of the
channel. It is a solution to a boundary value problem whereby the
channel’s extrema are subject to spin injection and spin pumping or
free spin boundary conditions. As a result, these
solutions exhibit peculiar characteristics of technological relevance,
namely~\cite{Iacocca2017d}: the spin injection threshold is
proportional to the square root of the in-plane anisotropy field for
long channels and the homogeneous frequency is inversely proportional
to damping and the channel’s length. For comparison, spin
waves~\cite{Stancil2009} excited on a homogeneous magnetization
background exhibit a spin injection threshold that is proportional to
damping, a frequency proportional to both spin injection and the
magnet's internal field, and an exponential decay rate that is
proportional to damping. The exponential decay of spin waves imposes
the ultimate limitation on their propagation length and coherent
spin transport, although detection at micrometer length scales has
been achieved in low-damping materials such as YIG~\cite{Liu2018},
amorphous YIG~\cite{Wesenberg2017}, and haematite~\cite{Lebrun2018}.

The analytical predictions and characteristics of DEFs are promising
for long-distance spin transport. However, the required spin injection
has emerged as a practical barrier for their experimental
realization. In recent experimental studies, spin injection was
realized from quantum Hall edge states in antiferromagnetic
graphene~\cite{Stepanov2018} and the spin-Hall effect in
Pt~\cite{Yuan2018}. A recent numerical study proposes an alternative
spin-injection mechanism based on the spin-transfer torque
effect~\cite{Slonczewski1996,Berger1996}, which excites magnetization
precession~\cite{Iacocca2017d}. This method allows for large spin
injection magnitudes, breaking the weak injection assumption that has
been analytically assumed to
date~\cite{Sonin2010,Takei2014}. Signatures of distinct nonlinear,
dispersive dynamics exhibiting solitonic features were observed in
micromagnetic simulations that include non-local dipole
fields~\cite{Iacocca2017d}.  More recently, micromagnetic simulations
that incorporate spin-transfer torque along a confined, central strip
of a ferromagnet have similarly shown evidence of strongly nonlinear
features including a soliton nucleated at the injection site in the
large injection regime termed a soliton screened spin
superfluid~\cite{Schneider2018}.

While the numerical studies to date by a variety of groups
unambiguously demonstrate that long-range spin transport can in
principle be achieved with noncollinear magnetization states in
magnetic materials, an analysis that incorporates short-wavelength exchange dispersion and large-amplitude nonlinearities due to anisotropy---such as those
necessarily present for the existence of a soliton---as well as a description of the effect of damping on spin flows is
lacking. Here, we provide a unified analytical framework in the
context of a dispersive hydrodynamic (DH) formulation of magnetization
dynamics~\cite{Iacocca2017,Iacocca2017b}. This formulation is an exact
transformation of the Landau-Lifshitz equation and,
therefore, captures the essential physics that are relevant to
describe fully nonlinear, noncollinear magnetization states: exchange,
anisotropy, and damping.

The DH formulation gives rise to two equations of motion for a
longitudinal spin density and its associated fluid velocity that are
analogous to the Navier-Stokes' mass and momentum equations for a
compressible fluid~\cite{Iacocca2017,Iacocca2017b}. From a fluid
perspective, exchange, anisotropy, and damping give rise to
dispersion, nonlinearity, and viscosity, respectively. In contrast to
typical fluids, the equivalent magnetic fluid exhibits a non-conserved
density, i.e., the mass can be lost. Therefore, noncollinear
magnetization states---DEFs---can be interpreted as forced fluid flows
that compensate the density and viscous losses manifesting in a
profile that balances dispersion and nonlinearity. 
 
In this paper, we find that DEFs are generally characterized by a
nonlinear profile in both density and fluid velocity. In the weak spin
injection regime, the DH equations reduce to the forced diffusion
equation and lead to a linear DEF solution that is equivalent to a
spin superfluid~\cite{Sonin2010,Takei2014}. Using boundary layer
theory in the strong spin injection regime, we find a novel dynamical
state characterized by the nucleation of a stationary soliton at the
injection site that smoothly transitions into a nonlinear DEF. We term
this dynamical solution as a contact soliton DEF or CS-DEF, which is
an analytical representation of the numerically identified soliton screened spin
superfluid~\cite{Schneider2018}. From a hydrodynamic perspective, the
soliton nucleated at the injection site occurs precisely when the
injection crosses the subsonic to supersonic flow boundary, equivalent
to the Landau criterion~\cite{Iacocca2017,Iacocca2017b}.  Moreover,
transition between a DEF and a CS-DEF corresponds to the maximum
precessional frequency achieved by spin injection, setting an upper
bound to the efficiency of DEF-mediated spin transport.  Thus, further
spin injection enhances the coherent, superfluid-like soliton at the
expense of larger spin transport, which is in sharp contrast to
classical fluids where strong channel flows are subject to drag at the
boundaries that, above a critical Reynolds number, develop into an
incoherent, turbulent state~\cite{avila_onset_2011}.

The presented results pertain to an ideal geometry whereby the
  magnetic material is defect-free and the boundaries are perfect
  spin-current sources and drains. Deviations from these conditions
  may result in qualitative changes to the presented solutions,
  including instabilities. Defects in the magnetic material can result
  in magnetic topological defects that destabilize the DEFs, e.g.,
  vortex-antivortex pairs~\cite{Iacocca2017b} or
  phase-slips~\cite{Sonin2010,Kim2016,Kim2016b}. Non-ideal boundaries
  can be incorporated by utilizing mixed (Robin) boundary conditions
  from a circuit formalism that includes spin
  pumping~\cite{Takei2014}. In the case of strong injection, recent
  numerical results suggest that such boundaries can induce an
  instability in the DEF to CS-DEF crossover
  region~\cite{Schneider2018}. Our results aim to provide the
  analytical basis to further study these effects in more detail.

Our analytical study also indicates that, for the physically relevant
case of magnetic materials with low damping, DEFs can be interpreted
as an adiabatic spatial evolution of conservative dynamic solutions,
previously termed uniform hydrodynamic states
(UHSs)~\cite{Iacocca2017} in order to highlight their non-dissipative, flowing character. DEF magnetization states sustained in channels
subject to subsonic spin injection conditions can be conveniently
represented as curves of constant frequency in the UHS phase space of
spin density and fluid velocity. From an applications perspective, the
fluid interpretation also lends itself to a circuit analogy, from
which we can define the current-voltage ($I$-$V$) characteristics of
the coherent states studied here. Micromagnetic simulations support
the analytical results even in the presence of in-plane anisotropy and
non-local dipole fields in a thin film.

The remainder of the paper is organized as follows. In Sec.~II, we
summarize the dispersive hydrodynamic formulation and main features of
uniform hydrodynamic states. In Sec.~III, we introduce the boundary
value problem that describes a channel subject to spin injection at
one extremum and derive analytical expressions for linear DEFs, DEFs,
and CS-DEFs. In the same section, we study the DEF to CS-DEF
transition in the context of a subsonic to supersonic flow
transition. In Sec.~IV, we establish that the hydrodynamic states
sustained in channels realize a nonlinear resistor in the hydraulic
analogy of electrical circuits. Micromagnetic simulations of nanowires
incorporating STT as a spin injection mechanism, in-plane
magnetocrystalline anisotropy, and non-local dipole fields are
discussed in Sec.~V. Finally, we provide our concluding remarks in
Sec.~VI.

\section{Dispersive hydrodynamic formulation and uniform hydrodynamic
  states}

Magnetization dynamics in a continuum approximation can be described
by the Landau-Lifshitz (LL) equation
\begin{subequations}
\begin{eqnarray}
\label{eq:1}
  \partial_t\mathbf{m} &=& -\mathbf{m}\times\mathbf{h}_\mathrm{eff} - \alpha\mathbf{m}\times\mathbf{m}\times\mathbf{h}_\mathrm{eff},\\
\label{eq:2}
  \mathbf{h}_\mathrm{eff} &=& \underbrace{\Delta\mathbf{m}}_\text{exchange} - \underbrace{m_z\hat{\mathbf{z}}}_\text{local dipole},
\end{eqnarray}
\end{subequations}
where $\mathbf{m}=(m_x,m_y,m_z)$ is the magnetization vector normalized to the saturation magnetization $M_s$, $\alpha$ is the phenomenological Gilbert damping parameter, and $\mathbf{h}_\mathrm{eff}$ is an effective field, normalized by $M_s$, that incorporates exchange and local (zero-thickness) dipole field as a minimal model for dispersion and nonlinearity, respectively. The dimensionless form of Eq.~\eqref{eq:1} is achieved by scaling time by $|\gamma|\mu_0M_s$ and space by $\lambda_\mathrm{ex}^{-1}$, where $\gamma$ is the gyromagnetic ratio, $\mu_0$ is the vacuum permeability, and $\lambda_\mathrm{ex}$ is the exchange length. A dispersive hydrodynamic representation of Eqs.~\eqref{eq:1} and \eqref{eq:2} can be achieved by mapping the magnetization vector into hydrodynamic variables~\cite{Iacocca2017,Iacocca2017b,Iacocca2017d}, namely, a longitudinal spin density $n=m_z$ and a fluid velocity $\mathbf{u}=-\nabla\Phi=-\nabla\arctan{(m_y/m_x)}$. In this work, we are interested in effectively one-dimensional dynamics along a channel whose length is oriented in the $\hat{\mathbf{x}}$ direction. Therefore, the fluid velocity can be written as a scalar quantity $u=\mathbf{u}\cdot\hat{\mathbf{x}}$ and the spatial derivatives taken only along $\hat{\mathbf{x}}$. The resulting dispersive hydrodynamic equations are
\begin{subequations}
\label{eq:3}
\begin{eqnarray}
  \label{eq:31}
    \partial_tn &=& (1+\alpha^2)\partial_x\left[(1-n^2)u\right] + \alpha(1-n^2)\partial_t\Phi,\\ 
  \label{eq:32}
    \partial_t\Phi &=& -(1-u^2)n+\frac{\partial_{xx}n}{1-n^2}+\frac{n (\partial_xn)^2}{(1-n^2)^2}\nonumber\\&&-\frac{\alpha}{1-n^2}\partial_x\left[(1-n^2)u\right].
\end{eqnarray}
\end{subequations}

The simplest solutions to Eq.~\eqref{eq:31} and \eqref{eq:32} are spin-density waves (SDWs). These are static ($\partial_t\Phi=0$), textured magnetization states parametrized by a constant density and fluid velocity, $(n_0,u_0)$. SDWs are magnetization states that support dissipationless spin transport~\cite{Konig2001}. A dynamic SDW can only be obtained as a transient state or in the conservative limit, where $\alpha=0$ and $\partial_t\Phi\neq0$. We refer to this state as a uniform hydrodynamic state (UHS). For both SDWs and UHSs, the density is limited by its deviation from the magnetization's unit sphere poles ($n = \pm1$ corresponds to vacuum) while the fluid velocity is an unbounded quantity. However, it was shown in Ref.~\onlinecite{Iacocca2017} that modulational instability~\cite{Zakharov2009} (the exponential growth of perturbations) ensues when $|u_0|>1$, i.e., for SDWs and UHSs with sub-exchange length, in-plane magnetization rotation wavelengths. Therefore, modulationally stable SDWs and UHSs are defined in the phase space spanned by $|n_0|<1$ and $|u_0|<1$. UHSs exhibit a precessional frequency given by
\begin{equation}
\label{eq:4}
  \Omega_0 = \partial_t\Phi = -\left(1-u_0^2\right)n_0,
\end{equation}
obtained directly from Eq.~\eqref{eq:32}. The negative sign of the frequency for $n_0>0$ indicates that the precession is clockwise about the $\hat{\mathbf{z}}$ direction.

\begin{figure}[t]
\centering \includegraphics[trim={0in 0in 0in 0in}, clip, width=3.5in]{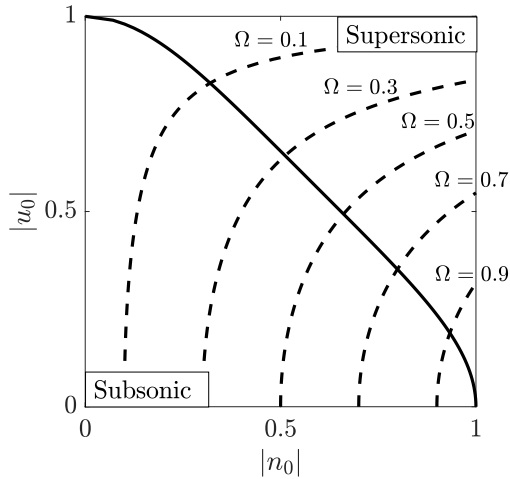}
\caption{ \label{fig0} UHS phase space for density $|n_0|\leq1$ and fluid velocity $|u_0|\leq1$. The sonic curve that separates the subsonic and supersonic regions is shown by a solid black curve. The dashed black curves represent isofrequency contours, labeled by the corresponding frequency.}
\end{figure}
It is important to emphasize that UHSs are dynamic, textured magnetization states. This is markedly different from small-amplitude perturbations about a homogeneous state that are typically associated with spin waves. Interestingly, UHSs support small-amplitude perturbations that exhibit a dispersion relation that is non-reciprocal for $n\neq0$~\cite{Iacocca2017,Iacocca2017b}. This nonreciprocity leads to conditions where long-wavelength perturbations can propagate in either two directions or one direction with respect to the UHS fluid velocity $u_0$ and can be hydrodynamically interpreted as subsonic or supersonic flow, respectively. The transition between subsonic and supersonic flow is known as the sonic curve. For UHSs, the sonic curve is given by
\begin{equation}
\label{eq:5}
  |u_0| = \sqrt{\frac{1-n_0^2}{1+3n_0^2}},
\end{equation}
and it is shown in Fig.~\ref{fig0} by a solid black curve in the UHS phase space.  Equation~\eqref{eq:5} is formally equivalent to the Landau criterion for superfluidity in the limit of perpendicularly magnetized easy-plane ferromagnets~\cite{Iacocca2017b} and for linear DEFs or spin superfluids~\cite{Sonin2018}. Isofrequency contours determined from Eq.~\eqref{eq:4} are shown by dashed black curves. As we will demonstrate below, the UHS phase space provides information regarding the form of dynamic magnetization states in ferromagnetic channels sustained by spin injection.

\section{Boundary value problem for easy-plane ferromagnetic channels}

The steady magnetization states sustained by spin injection can be
analytically obtained by solving Eqs.~\eqref{eq:31} and \eqref{eq:32}
subject to appropriate boundary conditions (BCs). For this, we
consider a channel of length $L$ and introduce spin injection at $x=0$
and free spin boundary conditions at $x=L$. For simplicity, we
disregard spin pumping~\cite{Takei2014}, but our analysis is sufficiently general that more complex BCs that incorporate metal / magnetic interfacial effects could be studied in a similar manner.

We seek steady, precessional solutions to
\begin{subequations}
\begin{eqnarray}
  \label{eq:33}
    0 &=& (1+\alpha^2)\frac{d}{dx}\left[(1-n^2)u\right]+\alpha(1-n^2)\Omega,\\ 
  \label{eq:34}
    \Omega &=& -(1-u^2)n+\frac{1}{1-n^2}\frac{d^2n}{dx^2}\\&&+\frac{n}{(1-n^2)^2}\left(\frac{dn}{dx}\right)^2-\frac{\alpha}{1-n^2}\frac{d}{dx}\left[(1-n^2)u\right].\nonumber
\end{eqnarray}
\end{subequations}
with BCs
\begin{subequations}
  \label{eq:18}
\begin{eqnarray}
\label{eq:61}
  \frac{dn}{dx}(0)=0, &\quad& \frac{dn}{dx}(L)=0,\\
\label{eq:62}
	u(0)=\bar{u}, &\quad& u(L)=0,
\end{eqnarray}
\end{subequations}
where $\bar{u}$ is proportional to the injected spin
current~\cite{Iacocca2017d}.  These boundary conditions are enforced
upon $n=n(x)$, $u=u(x)$ by introducing the homogeneous precessional
frequency $\Omega=\partial_t\Phi$. Below, we find solutions of this
boundary value problem (BVP) with nonlinearity, dispersion, and
damping. A more detailed, mathematical analysis leading to these approximate solutions is provided in the Appendices.

\subsection{Linear DEFs}

We begin our analysis by revisiting the weak spin injection regime
$0 < |\bar{u}|\ll \min(1,\alpha L)$, first presented in
\cite{Sonin2010,Takei2014}. For this, we assume that $u$ is small and $n$
is constant in Eqs.~\eqref{eq:33} and
\eqref{eq:34}, so that the linearized equations are
\begin{subequations}
  \label{eq:new1}
  \begin{eqnarray}
    \label{eq:new11}
    \alpha\tilde{\Omega} &=& -\frac{du}{dx},\\ 
    \label{eq:new12}
    \tilde{\Omega} &=& -n,
  \end{eqnarray}
\end{subequations}
where $\tilde{\Omega}=\Omega/(1+\alpha^2)$.

Noting that $u=-\partial_x\Phi$ and $\Omega=\partial_t\Phi$, we can
rewrite Eqs.~\eqref{eq:new1} as the diffusion equation
\begin{equation}
\label{eq:new2}
  \frac{\alpha}{1+\alpha^2}\partial_t{\Phi} = \partial_{xx}\Phi,
\end{equation}
subject to the boundary conditions
\begin{equation}
\label{eq:new3}
  \partial_x\Phi(0)=-\bar{u}, \quad \partial_x\Phi(L) = 0.\\
\end{equation}

For weak damping, $1+\alpha^2\sim1$, Eq.~\eqref{eq:new2} is the
linearized hydrodynamic diffusion equation for easy-plane ferromagnets from
previous studies~\cite{Sonin2010,Takei2014}. By direct integration,
Eq.~\eqref{eq:new2}, subject to Eq.~\eqref{eq:new3}, exhibits the linear
DEF solution
\begin{equation}
  \label{eq:def}
  u_{l\mathrm{DEF}} =
  \bar{u} (1-\frac{x}{L}), \quad\tilde{\Omega}_{l\mathrm{DEF}} =
  -n_{l\mathrm{DEF}} = \frac{\bar{u}}{\alpha L}
\end{equation}
that exhibits a linear decay profile in the fluid velocity, which
corresponds to the algebraic diffusion of spin current across the
channel.  Importantly, this approximate solution exhibits a spatially
homogeneous frequency and density. with no assumptions on the magnitudes of nonzero damping nor the channel length $L$. See Appendix~\ref{sec:linear-def-solution} for additional details.
\begin{figure*}[t]
\centering \includegraphics[trim={1in .5in 0.5in 0in}, clip, width=7in]{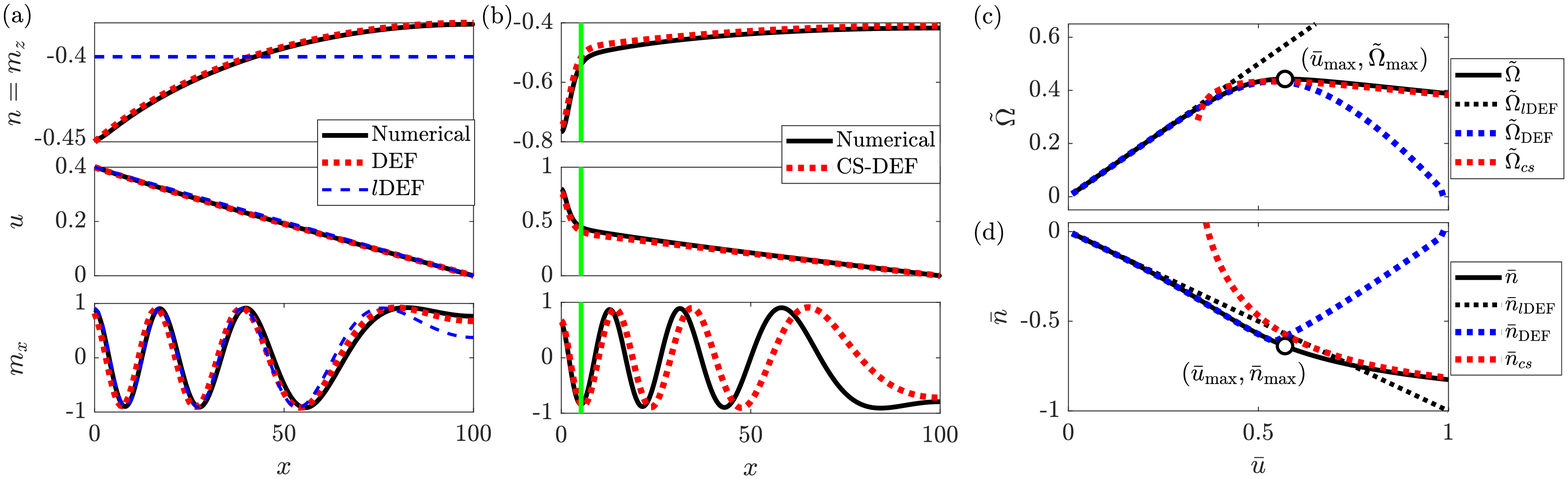}
\caption{ \label{fig1} (color online) Magnetization states in a channel of length $L=100$ and $\alpha=0.01$ subject to the injection $\bar{u}$ at the left edge, $x=0$. In (a) and (b), the panels represent the density $n$, fluid velocity $u$, and $m_x$ magnetization vectorial component at an instant of time. (a) For the injection $\bar{u}=0.4$, the numerical solution shown by solid black curves is in excellent agreement with a DEF shown by dashed red curves. For comparison, the corresponding linear DEF solution is shown by dashed blue curves. (b) For the injection $\bar{u}=0.8$, the numerical solution shown by solid black curves is in good agreement with a CS-DEF shown by dashed red curves.  The solid green line indicates the boundary layer length where the soliton is established. (c) Precessional frequency as a function of injection for a linear DEF (dashed black line), numerical solution of the BVP (solid black curve), DEF (dashed blue curve), and CS-DEF (dashed red curve). The numerical maximum $\tilde{\Omega}_\mathrm{max}=0.44$ is found at $\bar{u}=0.57$. (d) Density at the injection site, $\bar{n}=n(0)$, using the same color codes as (c). }
\end{figure*}

It is important to emphasize that damping plays a fundamental role in
the stabilization of the linear DEF solution.  It is for this reason
that we refer to the solution as a dissipative exchange flow.  In
fact, in the conservative case where $\alpha=0$, the solution to
Eq.~\eqref{eq:new11} ($u = $ const) cannot satisfy both boundary
conditions \eqref{eq:new3}.

\subsection{Nonlinear DEFs}

We now consider nonlinear but spatially smooth solutions, i.e., slowly
varying relative to the exchange length for a long channel $L\gg1$. Consequently, the dispersive
terms in Eq.~\eqref{eq:32} can be neglected (both $d^2n/dx^2$ and
$(dn/dx)^2$). Upon simple algebraic manipulation, Eqs.~\eqref{eq:33}
and \eqref{eq:34} reduce to
\begin{subequations}
\label{eq:7}
\begin{eqnarray}
  \label{eq:71}
    \alpha(1-n^2)\tilde{\Omega} &=& -\frac{d}{dx}\left[(1-n^2)u\right],\\ 
  \label{eq:72}
    \tilde{\Omega} &=& -(1-u^2)n .
\end{eqnarray}
\end{subequations}
Inserting $n$ from Eq.~\eqref{eq:72} into \eqref{eq:71} leads, after some algebra, to the differential equation
\begin{equation}
\label{eq:73}
  \alpha\tilde{\Omega} = \frac{du}{dx}\left[\frac{(\alpha\tilde{\Omega}u)^2}{(1-u^2)(u^4-2u^2+1-\tilde{\Omega}^2)}-1\right],
\end{equation}
that relates the fluid velocity to the precessional frequency. By integration, we obtain an implicit equation for the fluid velocity (see Appendix~\ref{app:def}) 
\begin{eqnarray}
\label{eq:8}
  \alpha L\tilde{\Omega}_\mathrm{DEF}\left(1-\frac{x}{L}\right)&=&u_\mathrm{DEF}+4\tanh^{-1}{(u_\mathrm{DEF})}\nonumber\\&-&2\mathcal{N}^-(u_\mathrm{DEF},\tilde{\Omega}_\mathrm{DEF})\nonumber\\&-&2\mathcal{N}^+(u_\mathrm{DEF},\tilde{\Omega}_\mathrm{DEF}),
\end{eqnarray}
where
\begin{equation}
  \label{eq:81}
  \mathcal{N}^\pm (\kappa,\omega) = \sqrt{1\pm\omega}
  \tanh^{-1}
  \left(\frac{\kappa}{\sqrt{1\pm\omega}} \right).
\end{equation}

The precessional frequency is obtained by evaluating Eq.~\eqref{eq:8} at $x=0$ where $u_\mathrm{DEF}(0)=\bar{u}$, implying the equation for the DEF's frequency
\begin{eqnarray}
\label{eq:82}
  \alpha L\tilde{\Omega}_\mathrm{DEF} &=& \bar{u}+4\tanh^{-1}{(\bar{u})}\nonumber\\&-&2\left[\mathcal{N}^-(\bar{u},\tilde{\Omega}_\mathrm{DEF})+\mathcal{N}^+(\bar{u},\tilde{\Omega}_\mathrm{DEF})\right],
\end{eqnarray}
while the density is obtained directly from Eq.~\eqref{eq:72} as
\begin{equation}
\label{eq:83}
  n_\mathrm{DEF} = -\frac{\tilde{\Omega}_\mathrm{DEF}}{1-u_\mathrm{DEF}^2}.
\end{equation}

Equations~\eqref{eq:8}, \eqref{eq:82}, and \eqref{eq:83} indicate that the DEF's spatial profile is, in general, nonlinear and the frequency is a nonlinear function of the spin injection $\bar{u}$. A numerical solution for a nonlinear DEF is shown by dashed red curves in Fig.~\ref{fig1}(a) for the injection $\bar{u}=0.4$, a channel of length $L=100$, and $\alpha=0.01$. The top and center panels show the hydrodynamic variables $n(x)$ and $u(x)$, respectively, while the bottom panel shows the $\hat{\mathbf{x}}$ magnetization component, $m_x(x,t)=\sqrt{1-n(x)^2}\cos{\Phi(x)}$ at a given instant of time (recall that $\partial_t\Phi\neq0$). Excellent agreement is obtained between the analytical solution and the numerical solution of the full BVP in Eqs.~\eqref{eq:31}, \eqref{eq:32}, \eqref{eq:61}, and \eqref{eq:62}, shown by solid black curves. The BVP is numerically solved by a collocation method (MATLAB's bvp5c).

An important consequence of the DEF nonlinear profile is the concomitant precessional frequency that is a nonlinear function of the injection, $\bar{u}$, shown by a dashed blue curve in Fig.~\ref{fig1}(c). The frequency obtained by solving the full BVP is shown by a solid black curve. Excellent agreement with Eq.~\eqref{eq:82} is found up to the maximum frequency $\tilde{\Omega}_\mathrm{max}=0.44$ at $\bar{u}_\mathrm{max}=0.57$, indicated by a black circle. For $\bar{u}>\bar{u}_\mathrm{max}$, the nonlinear solution no longer describes the frequency dependence. The density at the injection site, equivalent to the magnetization tilt due to spin injection, is shown in Fig.~\ref{fig1}(d). Similar to the precessional frequency, a good quantitative agreement between the numerical solution (solid black curve) and the DEF solution (dashed blue curve) is observed up to $\bar{u}_\mathrm{max}=0.57$, where $\bar{n}_\mathrm{max}=-0.64$. As we show below, these qualitative changes indicate the initiation of supersonic flow and of a stationary soliton. 

The linear DEF solution can be obtained from the nonlinear DEF solution in
the weak injection regime. For this, we note that
$\tanh^{-1}(\kappa)\approx\kappa$ and
$\mathcal{N}^\pm(\kappa,\omega)\approx\kappa$ for small $\kappa$. Introducing these
approximations in Eqs.~\eqref{eq:8}, \eqref{eq:82}, and \eqref{eq:83}
leads to Eq.~\eqref{eq:def}.

The linear DEF approximation is shown by dashed blue curves in Fig.~\ref{fig1}(a) for the same parameters as the DEF and numerical solutions. It is interesting that while the difference between the linear and nonlinear spatial profiles for the fluid velocity (middle panel) is imperceptible, the density in a linear approximation does not conform to the spatial profile. A consequence is that the linear DEF frequency tuneability is likewise a linear function of injection and quantitatively agrees with the nonlinear solution up to $\bar{u}\approx0.3$ for $L=100$ and $\alpha=0.01$, shown in Fig.~\ref{fig1}(c) by a dashed black line. 

\subsection{Contact soliton DEFs}

The qualitative change in the frequency dependence observed in Fig.~\ref{fig1}(c) is an indication that the inclusion of nonlinearity and lowest order dispersion are not sufficient to describe DEF solutions sustained at an arbitrary injection strength. In such a regime, {higher order} dispersive terms must be taken into account in Eqs.~\eqref{eq:33} and \eqref{eq:34}. An analytical methodology for this task is boundary layer theory~\cite{Bender1999}. This method allows one to separate the system into regimes dominated by different physics that can be asymptotically matched. Below we outline the most important features and results obtained from the calculation. {Details} can be found in Appendix~\ref{app:bl}.

For Eqs.~\eqref{eq:33} and \eqref{eq:34} subject to the BCs~\eqref{eq:61} and \eqref{eq:62}, it is possible to {identify} two regimes. Close to the left edge subject to strong injection, the spatial profile of the solution can vary rapidly. In other words, we assume that dispersion dominates over damping. Asymptotically, this is equivalent to {an expansion with small damping while considering short spatial variations}, as discussed in the Appendix. We refer to this region as the inner region. Far from the left edge, we {assume} that the spatial profile of the solution varies slowly, so that damping dominates over dispersion. We refer to this region as the outer region. {A matching condition} is invoked to obtain a smooth solution across both regions. Mathematically, this is achieved by introducing BCs for the inner region
\begin{subequations}
\begin{eqnarray}
\label{eq:63}
  \frac{d}{dx}n_{in}(0)=0, &\quad& \lim_{x\rightarrow\infty}{n_{in}}(x)=n_\infty,\\
\label{eq:64}
	u_{in}(0)=\bar{u}, &\quad& \lim_{x\rightarrow\infty}{u_{in}}(x)=u_\infty,
\end{eqnarray}
\end{subequations}
and the outer region,
\begin{subequations}
\begin{eqnarray}
\label{eq:65}
  \lim_{x\rightarrow0}{n_{out}}(x)=n_\infty, &\quad& \frac{d}{dx}n_{out}(L)=0,\\
\label{eq:66}
	\lim_{x\rightarrow0}{u_{out}}(x)=u_\infty, &\quad& u_{out}(L)=0,
\end{eqnarray}
\end{subequations}
where $n_\infty$ and $u_\infty$ are matching {parameters} to be determined.

The equations of motion for the inner region are dominated by dispersion so that the dissipative terms are neglected
\begin{subequations}
\label{eq:10}
\begin{eqnarray}
  \label{eq:101}
    0 &=& \frac{d}{dx}\left[(1-n^2)u\right],\\ 
  \label{eq:102}
    \tilde{\Omega} &=& -(1-u^2)n+\frac{1}{1-n^2}\frac{d^2n}{dx^2}+\frac{n}{(1-n^2)^2}\left(\frac{dn}{dx}\right)^2.
\end{eqnarray}
\end{subequations}

The solution of this system of differential equations involves a series of steps detailed in Appendix~\ref{app:bl}. Ultimately, Eqs.~\eqref{eq:101} and \eqref{eq:102} can be integrated to obtain the soliton solution, e.g., see Ref.~\onlinecite{Congy2016}
\begin{subequations}
\label{eq:11}
\begin{eqnarray}
  \label{eq:111}
    n_{in} &=& \frac{a\nu_1\mathrm{tanh}^2{(\theta x)}+\nu_2(n_\infty-a)}{a\mathrm{tanh}^2{(\theta x)}+\nu_2},\\
  \label{eq:112}
    u_{in} &=& u_\infty\frac{1-n_\infty^2}{1-n_{in}^2},\\
	\label{eq:113}
		\tilde{\Omega}_{in} &=& -n_\infty(1-u_\infty^2),
\end{eqnarray}
\end{subequations}
with two free parameters: $n_\infty$, $u_\infty$. The coefficients $\nu_1$, $\nu_2$, $\theta$, and $a$ are given in Appendix~\ref{app:bl} and all BCs in Eqs.~\eqref{eq:63} and \eqref{eq:64} were used. In other words, Eqs.~\eqref{eq:111} and \eqref{eq:112} describe, respectively, solitons of density amplitude $a$ on a nonzero density background $n_\infty$ and fluid velocity background $u_\infty$.

In contrast, the slowly varying outer region is dominated by damping, leading to Eqs.~\eqref{eq:71} and \eqref{eq:72} with DEF solutions given by Eqs.~\eqref{eq:8} and \eqref{eq:83} we term $u_{out}$ and $n_{out}$, respectively. We note that this solution is obtained by evaluating the BCs of Eqs.~\eqref{eq:65} and \eqref{eq:66} at $x=L$, yielding a two-parameter family of solutions
\begin{subequations}
\label{eq:11new}
\begin{eqnarray}
  \label{eq:111ne}
    n_{out} = -\frac{\tilde{\Omega}_{out}}{1-u_{out}^2},\\
  \label{eq:112new}
    \alpha L\tilde{\Omega}_{out}\left(1-\frac{x}{L}\right)&=&u_{out}+4\tanh^{-1}{(u_{out})}\nonumber\\&-&2\left[\mathcal{N}^-(u_{out})+\mathcal{N}^+(u_{out})\right].
\end{eqnarray}
\end{subequations}

To apply boundary layer theory, the inner and outer solutions must asymptotically match and exhibit a single precessional frequency $\tilde{\Omega}_{cs}=\tilde{\Omega}_{in}=\tilde{\Omega}_{out}$. For the left edge of the channel subject to spin injection, we evaluate the inner region solution, Eqs.~\eqref{eq:111} and \eqref{eq:112} at $x=0$, to obtain
\begin{equation}
\label{eq:131}
  \bar{u} = u_\infty\frac{1-n_\infty^2}{1-(n_\infty-a)^2}.
\end{equation}

Then, we evaluate the matching conditions {applied to the outer solution}, Eqs.~\eqref{eq:65} and
\eqref{eq:66}, by {evaluating} Eqs.~\eqref{eq:111ne}, \eqref{eq:112new} {at} $x
= 0$ and {identifying} $u_{out}(0)= u_\infty$ and $n_{out}(0)=n_\infty$.

We now have all the ingredients to construct {a} uniformly valid
solution along the length $L$ of the channel. Such a solution can be
written as
\begin{subequations}
\label{eq:12}
\begin{eqnarray}
  \label{eq:121}
	  u_{cs}(x) &=& u_{in}(x) + u_{\mathrm{DEF}}(x) - u_\infty,\\
	\label{eq:122}
    n_{cs}(x) &=& n_{in}(x) + n_{\mathrm{DEF}}(x) - n_\infty,
\end{eqnarray}
\end{subequations}
{which} describes a soliton located at the injection site smoothly
connected to a nonlinear DEF. 
We call this solution a contact soliton {dissipative exchange flow} (CS-DEF).

A CS-DEF is shown by dashed red curves in Fig.~\ref{fig1}(b) for {the} injection $\bar{u}=0.8$, a channel of length $L=100$, and $\alpha=0.01$. The numerical solution of the full BVP is shown by solid black curves and it is in excellent quantitative agreement to the boundary layer approach. The frequency dependence {on} the injection $\bar{u}$ is shown by a dashed red curve in Fig.~\ref{fig1}(c). In contrast to the DEF frequency tuneability, the CS-DEF precessional frequency is {decreasing with $\bar{u}$}. Additionally, we observe that the numerically obtained frequency tuneability, solid black line, approaches the CS-DEF frequency above $\bar{u}_\mathrm{max}$. A similar behavior is observed for the density at the injection site, shown in Fig.~\ref{fig1}(d) by the dashed red curve. These observations indicate that the full {solution} profile as a function of injection $\bar{u}$ transitions from a DEF into a CS-DEF. In the following section, we investigate this transition and its hydrodynamic interpretation.

Qualitatively, CS-DEFs are similar to the soliton screened spin superfluid recently {calculated} in micromagnetic simulations~\cite{Schneider2018}. An important difference is that our free-spin boundary conditions model a perfect spin sink so that magnon reflections are inhibited.

\subsection{DEF to CS-DEF transition}
\label{sec:def-cs-def}

In the previous section, a transition from a DEF into a CS-DEF was evidenced {by} a qualitative change {in} the frequency tuneability to injection. In particular, it is observed in Fig.~\ref{fig1}(c) that the full numerical solution (solid black curve) approaches the DEF and CS-DEF frequency tuneabilities in the small and large injection {limits}, respectively. Whereas a first-order transition is not observed, it is insightful to find an analytical expression for a practical observable, such as the maximum precessional frequency, $\tilde{\Omega}_\mathrm{max}$. For this, we can utilize the implicit equation for a DEF fluid velocity profile, Eq.~\eqref{eq:82}, to take the derivative with respect to $u$ and equate {$d/d\bar{u}(\tilde{\Omega}_\mathrm{DEF})=0$}. Because Eq.~\eqref{eq:82} is implicit, the maximum frequency will be an implicit equation as well. Utilizing Eq.~\eqref{eq:83}, we can {eliminate} $\tilde{\Omega}_\mathrm{DEF}$ and, after some algebra, we {obtain the} injection {at maximum frequency, $\bar{u}_\mathrm{max}$, that depends on the} input density {at maximum frequency}, $\bar{n}_\mathrm{max}$, {according to}
\begin{equation}
\label{eq:9}
  |\bar{u}_\mathrm{max}| = \sqrt{\frac{1-\bar{n}_\mathrm{max}^2}{1+3\bar{n}_\mathrm{max}^2}}.
\end{equation}
Interestingly, this is {precisely} the sonic curve, Eq.~\eqref{eq:4}. This {relation} is a central result of this work.

\begin{figure}[t]
\centering \includegraphics[trim={0in 0.8in 0in .5in}, clip, width=3.5in]{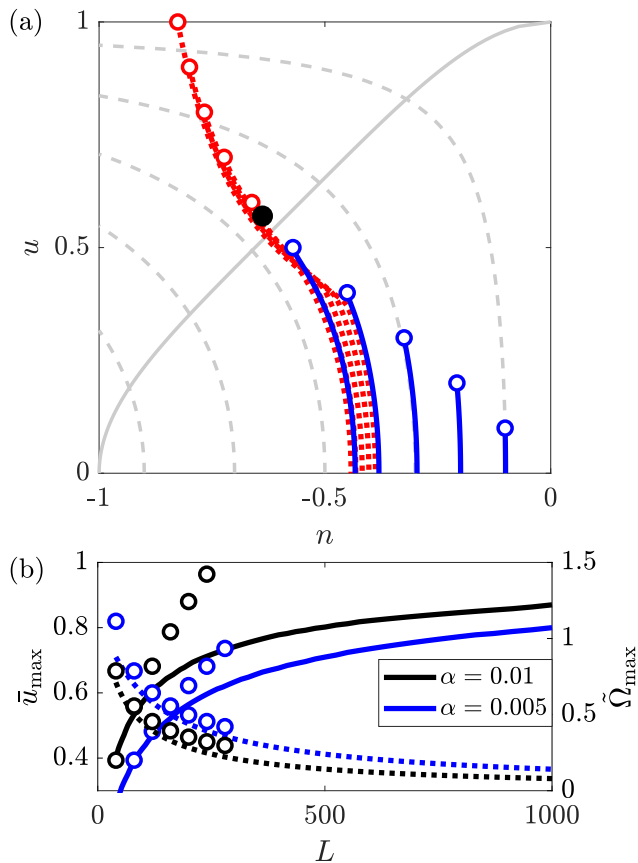}
\caption{ \label{fig2} (color online) (a) DEFs (solid blue curves) and CS-DEFs (dashed red curves) represented in the UHS phase space. The sonic curve and isofrequency contours are shown by a solid and dashed gray curves, respectively. The DEFs {lie on} the isofrequency contours, in agreement with an adiabatic interpretation. CS-DEFs behave markedly different when the parameters are in the supersonic regime. {The density and injection at the frequency maximum for $L=100$, $(\bar{n}_\mathrm{max},\bar{u}_\mathrm{max})$, is shown by a black circle.} (b) injection (left axis, solid curves) and frequency (right axis, dashed curves) at which a DEF transitions into a CS-DEF as a function of the channel length $L$ and setting $\alpha=0.01$ {(black) and $\alpha=0.005$ (blue). Analytical estimates obtained from an asymptotic expansion in $\bar{u}$ of the nonlinear DEF solution are shown by circles with the color code described above}.}
\end{figure}
There are three physical implications of Eq.~\eqref{eq:9}. First, the relation bounds the phase space for DEFs to the UHS subsonic regime{, below the solid curve in Fig.~\ref{fig0}}. Second, it suggests that DEFs can be interpreted as {the} adiabatic {spatial} evolution through a family of UHSs parametrized by spatially-dependent densities and fluid velocities. An adiabatic interpretation is valid as long as $\alpha\ll1$, which is physically true for magnetic materials of interest. Third, exceeding $\bar{u}_\mathrm{max}$ implies supersonic flow and coincides with the development of a soliton at the injection site.

A consequence of the adiabatic interpretation of DEF solutions is that the solution's profiles can be visualized {within} UHS phase space. In Fig.~\ref{fig2}(a), we show numerical solutions of the BVP for $L=100$ and $\alpha=0.01$ by solid blue curves. The input conditions for each case are marked by blue circles. The solid and dashed gray curves represent the UHS sonic curve and isofrequency contours, respectively. We observe that the density and fluid velocity of several DEFs {lie on} UHS isofrequency contours. When the injection and its corresponding density enter the supersonic regime, CS-DEFs ensue and the adiabatic interpretation breaks down. Numerical solutions for CS-DEFs visualized in the UHS phase space are shown by dashed red curves in Fig.~\ref{fig2}(a) where the input conditions are marked by red circles. Close to the injection {site}, where the soliton is established, the profile does not follow the isofrequency contours. However, once the sonic curve is crossed, the profile transitions into that of a DEF and {spatially} evolves adiabatically {along an isofrequency contous in UHS phase space}.

From a hydrodynamic perspective, the UHS phase space visualization emphasizes a remarkable quality of CS-DEFs. In classical fluids, {high speed flow with} boundaries is subject to instabilities that result in turbulent flow, i.e., characteristic spatial scales become smaller downstream. Instead, the soliton established at the injection site is a coherent structure that expands the spatial scales to a slowly varying DEF, precluding turbulence and ultimately establishing a {slower} subsonic flow. This feature is possible at the expense of reducing the homogeneous precessional frequency and, consequently, the magnitude of spin currents pumped into a reservoir located, e.g., at the right edge of the channel. It must be noted that supersonic conditions close to the {left} edge of the channel make this region susceptible to instabilities via phase slips~\cite{Sonin2010} or vortex-antivortex pair creation~\cite{Iacocca2017b} at defect sites. {A} detailed study of CS-DEF instabilities as well as the conditions that trigger such instabilities {is a separate study}.

As discussed above, the distinction between DEFs and CS-DEFs from a hydrodynamic perspective can be linked to the flow conditions at the injection site. However, Eq.~\eqref{eq:9} is expressed as a function of $\bar{n}_\mathrm{max}$, which is an {a priori} unknown quantity that is determined by solving for a DEF. In other words, Eq.~\eqref{eq:9} cannot predict which isofrequency contour in Fig.~\ref{fig2}(a) will be followed by a DEF given only the injection $\bar{u}$. A practical consequence is that the actual maximum injection and precessional frequency will depend on $L$ and $\alpha$. By numerically solving the BVP as a function of $L$, we find the maximum injection $\bar{u}_\mathrm{max}$ and frequency $\tilde{\Omega}_\mathrm{max}$ shown, respectively, by solid and dashed curves in Fig.~\ref{fig2}(b) {for $\alpha=0.01$ (blue) and $\alpha=0.005$ (black). The density and injection at the frequency maximum for $L=100$, $(\bar{n}_\mathrm{max},\bar{u}_\mathrm{max})$, is shown by a black circle in Fig.~\ref{fig2}(a).} These results have a clear physical interpretation. For short channels, the problem limits to a local balance between injection and damping. Therefore, the energy introduced into the system is primarily invested in spin precession. In the opposite limit of long channels, the energy is mainly invested in establishing a DEF to compensate damping nonlocally and $\bar{u}_\mathrm{max}$ is large.

{Analytical expressions for for both $\bar{u}_\mathrm{max}$ and $\tilde{\Omega}_\mathrm{max}$ can be obtained from the asymptotic expansion in $\bar{u}$ of the nonlinear DEF solution, written in the Appendix~\ref{app:def}. Following the same procedure outline above, we obtain}
\begin{subequations}
\begin{eqnarray}
\label{eq:uder_new}
  \bar{u}_\mathrm{max} &\approx& \left(\frac{3}{20}\right)^{1/4}\sqrt{\alpha L} \approx 0.6223\sqrt{\alpha L},\\
\label{eq:omegader_new}
  \tilde{\Omega}_\mathrm{max} &\approx& \left(\frac{3}{20}\right)^{1/4}\frac{4}{5}\frac{1}{\sqrt{\alpha L}} \approx 0.4979\frac{1}{\sqrt{\alpha L}}.
\end{eqnarray}
\end{subequations}

{These solutions are valid for small $\alpha L$. Comparison to our analytical results are shown in Fig.~\ref{fig2}(b) with black and blue circles for, respectively, $\bar{u}_\mathrm{max}$ and $\tilde{\Omega}_\mathrm{max}$. For the typical small values of $\alpha$, good agreement is observed up to $L\approx 200$.}

We emphasize that neither {in-plane} anisotropy nor non-local dipole fields have been {included in the analysis}. For short channels, these fields will most likely change the easy axis direction, which {could} destroy the onset of magnetization textures. However, for long channels, it has been shown that such symmetry-breaking fields primarily introduce a threshold for the onset of DEFs~\cite{Iacocca2017d}. This implies that the large injections required to trigger a transition into a CS-DEF will be negligibly affected, as recently observed by simulations~\cite{Schneider2018}. In section V, we explore this transition by micromagnetic simulations in nanowires where the injection is parametrized by STT.

\subsection{Boundary layer {width}}

The CS-DEF solution presented in Eqs.~\eqref{eq:121} and \eqref{eq:122} {was} obtained by separating the problem {into} two distinct regions, inner and outer, followed by asymptotic matching. A relevant parameter to {identify} is the {width} of the {solitonic} inner region as a function of the injection $\bar{u}$.

The boundary layer {width} is linked to the soliton width, whose profile is given in Eq.~\eqref{eq:111}. Because solitons decay exponentially, its width can be estimated from the profile's half-width at half-maximum. We will use this metric to estimate the boundary layer {width}, $l$.
\begin{figure}[t]
\centering \includegraphics[trim={0in 0in 0in 0in}, clip, width=3.5in]{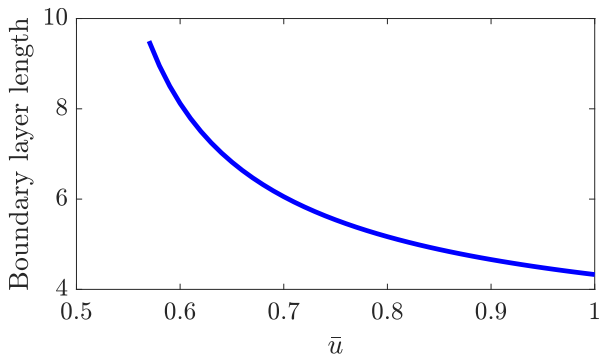}
\caption{ \label{fig:extra} (color online) Boundary layer {width} as a function of the injection strength $\bar{u}$ }
\end{figure}

The soliton solution Eq.~\eqref{eq:111} has an amplitude $a$ over a background $n_\infty$. Therefore, the half-width at half-maximum can be calculated by imposing {$n_{in}(x=l)=-a/2+n_\infty$}. After some algebra, we obtain the implicit equation for $l$
\begin{equation}
\label{eq:new_1}
  \tanh^2{(\theta l)} = \frac{\nu_2}{2(\nu_1-n_\infty)+a},
\end{equation}
that can be solved numerically as a function of $\bar{u}$ given the boundary and matching conditions~\eqref{eq:111ne}, \eqref{eq:112new}, and \eqref{eq:131}. Figure~\ref{fig:extra} {depicts} the boundary layer {width} as a function of $\bar{u}$ {larger than $\bar{u}_\mathrm{max}$}, where the CS-DEF solution occurs in {a} channel {of} length $L=100$. We observe that the boundary layer {width} decreases, i.e., the soliton becomes sharper, with injection strength. For the particular case of $\bar{u}=0.8$, the solution to Eq.~\eqref{eq:new_1} predicts a boundary layer {width} of $\approx 5$. This is shown by the vertical solid green line in Fig.~\ref{fig1}, in good agreement with both the numerical calculation and the analytical solution.

The boundary layer {width} of Fig.~\ref{fig:extra} is presented in units of exchange length, valid for easy-plane anisotropy materials. For Permalloy with a typical exchange length of $5$~nm, the boundary layer {width} lies between $22$~nm and $47$~nm in a channel of $500$~nm. For {parameters associated with} YIG~\cite{Schneider2018}, $A=3.5$~pJ/m and $M_s=130$~kA/m, the exchange length is $\approx18$~nm. This leads to a boundary layer {width} between $78$~nm and $172$~nm in a channel of $1.8$~$\mu$m.

\section{Electrical circuit analogy}

An alternative interpretation that captures the behavior of the channel subject to injection as a two-terminal device is the hydraulic analogy {to} electrical circuits. This analogy allows one to classify the DEFs and CS-DEFs in the context of electrical elements that provide building blocks to construct devices with a given functionality. For this, we define hydrodynamic quantities that are analogous to a voltage and a current, and from which the I-V characteristics of the device can be obtained.

In the {electric to} hydraulic analogy, a voltage maps to pressure difference. Using the hydrodynamic formulation of magnetization dynamics, the spatially-dependent pressure $P(x)$ was derived in Ref.~\onlinecite{Iacocca2017} as
\begin{equation}
\label{eq:pressure}
  2P(x) = [1+n(x)^2][1-|\mathbf{u}(x)|^2]-1,
\end{equation}
from which the pressure difference or voltage $V=P(x=L)-P(x=0)$ in a channel of length $L$ subject to BCs~\eqref{eq:62} is
\begin{equation}
\label{eq:voltage}
  V = \frac{1}{2}\left[(n_L^2-\bar{n}^2)+(1+\bar{n}^2)\bar{u}^2\right],
\end{equation}
where $n_L=n(x=L)$ and {$\bar{n}=n(x=0)$} are the densities at the channel's extrema.

\begin{figure}[t]
\centering \includegraphics[trim={0in 0in 0in 0in}, clip, width=3.5in]{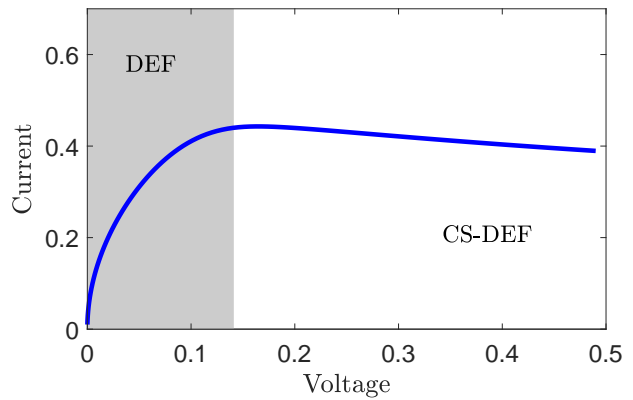}
\caption{ \label{fig:circuit} I-V characteristics for a channel of length $L=100$ and $\alpha=0.01$ subject to a spin injection $\bar{u}$ at $x=0$. The gray and white area indicate the regions where, respectively, DEFs and CS-DEFs are sustained. }
\end{figure}
{The} current $I$ is equivalent to {the} density {flux}. In the steady state modes studied here, the density {flux is $(1-n^2)\mathbf{u}$, whose magnitude} corresponds to the precessional frequency, Eq.~\eqref{eq:4}. Note that the precessional frequency is the only spatially-homogeneous quantity of both DEFs and CS-DEFs, just as a current is an equilibrium, constant quantity in electric circuits. Additionally, in the case of a neighboring spin reservoir, the precessional frequency is linearly dependent {on} the pumped spin current that can give rise to a transverse charge current by {the} inverse spin-Hall effect~\cite{Hoffmann2013}.

Using Eq.~\eqref{eq:voltage} and Eq.~\eqref{eq:4}, we numerically calculate the I-V characteristics shown in Fig.~\ref{fig:circuit} for a channel of length $L=100$ and $\alpha=0.01$. The gray and white areas indicate the sustenance of, respectively, a DEF or a CS-DEF. The I-V characteristic is nonlinear for all cases and its finite value indicates that both DEFs and CS-DEFs are \emph{resistive}. In other words, hydrodynamic states sustained in channels subject to spin injection can be classified as nonlinear resistors.

We note that in this representation, even the linear DEF solution Eq.~\eqref{eq:def}, results in a nonlinear I-V curve. In fact, the linear solution establishes a spatially constant density, so that {$n_L=\bar{n}$}. Additionally, {$|\bar{n}|\ll1$}, leading to a voltage given simply by $V=\bar{u}^2/2$. The precessional frequency is given in Eq.~\eqref{eq:def} so that $I=\bar{u}/(\alpha L)$. Therefore, the resistance is $R=V/I=\alpha L \bar{u}/2=\alpha L\sqrt{V/2}$.

A notable feature of the I-V curve is the change {in} slope from positive when a DEF is sustained to negative when a CS-DEF is sustained. This agrees with the frequency tuneability shown in Fig.~\ref{fig1}(c). In terms of the differential conductivity, dI$/$dV, this implies a positive {or} negative sign for, respectively, DEFs and CS-DEFs. While the I-V characteristic is positive everywhere, the negative differential conductivity of CS-DEFs implies that these states can potentially amplify oscillatory inputs.

\section{Micromagnetic simulations}

In this section, we explore the DEF solutions established in a nanowire by micromagnetic simulations including both non-local dipole fields and magnetocrystalline anisotropy. We utilize the GPU-based code mumax3~\cite{Vansteenkiste2014}. We consider material {parameters for} Py, namely, $M_s=790$~kA/m, exchange stiffness $A=10$~pJ/m, in-plane anisotropy field $H_A=400$~A/m, and $\alpha=0.01$. The corresponding exchange length {for} these parameters is $\lambda_\mathrm{ex}=5.05$~nm.

We simulate a nanowire of dimensions $512$~nm~$\times~100$~nm~$\times~1$~nm. Spin injection is achieved by STT acting on a $10$~nm~$\times~100$~nm contact located at the left extremum of the nanowire. Therefore, the nanowire length subject to spin injection is $502$~nm that corresponds to a dimensionless length of $L=99.4$. We use a symmetric STT with polarization $P=0.65$ and assume that the charge current is spin-polarized along the $\hat{\mathbf{z}}$ direction, e.g., by a magnetic material with perpendicular magnetic anisotropy~\cite{Bruno1989}. From a previous study~\cite{Iacocca2017d}, it was found that DEFs can be excited by STT in the presence of symmetry-breaking terms by charge current densities on the order or $10^{11}$~A/m$^2$. We numerically find a threshold of $\bar{J}=4\times10^{11}$~A/m$^2$. To explore the dynamical regimes discussed in Sec.~III, we vary the charge current density at the left contact, between $1\times10^{11}$~A/m$^2$ and $20\times10^{11}$~A/m$^2$ in steps of $1\times10^{11}$~A/m$^2$. The simulation was set to run for $20$~ns for each current, which was found to be sufficient to stabilize a steady state regime.
\begin{figure}[t]
\centering \includegraphics[trim={0in 1.in 0in .5in}, clip, width=3in]{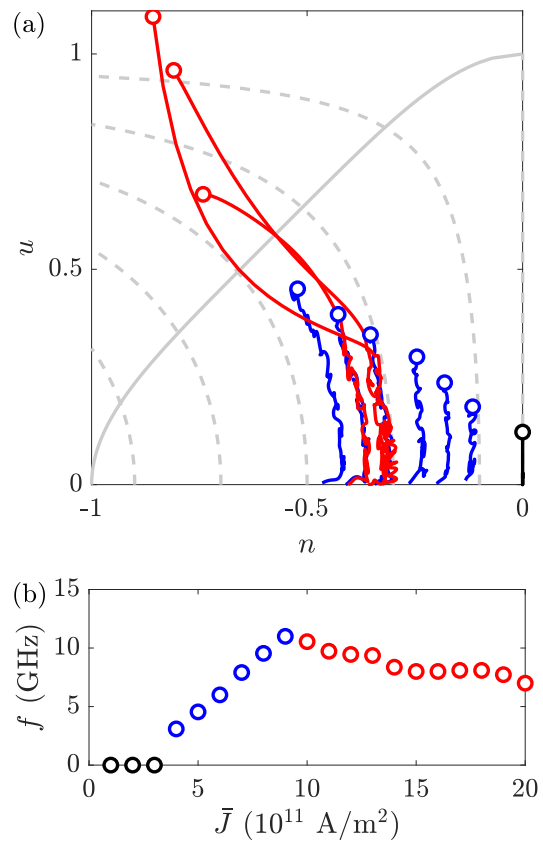}
\caption{ \label{fig5} (a) Magnetization states {shown in the UHS phase space} obtained from micromagnetic simulations of a Py nanowire of dimensions $512$~nm~$\times~100$~nm~$\times~1$~nm subject to STT at the left extremum. The injection conditions are identified by circles and the corresponding solutions are plotted as solid curves. Sub -threshold solutions are shown in black, DEFs in blue, and CS-DEFs in red. The solid and dashed gray curves represent the sonic curve and isofrequency contours for a UHS, respectively. (b) Frequency tuneability as a function of $\bar{J}$. }
\end{figure}

The results can be visualized in the UHS phase space shown in Fig.~\ref{fig5}(a). Because of the oscillations and transverse non-uniformity introduced by anisotropy and non-local dipole fields~\cite{Iacocca2017d}, respectively, we plot averaged densities and fluid velocities. The average is performed both in space across the width of the nanowire and in time for the range $15$~ns to $20$~ns. {To directly compare with the analytical results, we disregard the region subject to STT. In other words, the boundary conditions are determined just outside the region subject to STT in the nanowire and the nanowire’s right extremum.} A current density threshold for the stabilization of hydrodynamic states is observed. At sub-threshold current densities, a partial domain wall is formed at the injection site~\cite{Iacocca2017d}, evidenced by a solid black vertical line at $n=0$.
\begin{figure}[t]
\centering \includegraphics[trim={0in 0in 0in 0in}, clip, width=3.5in]{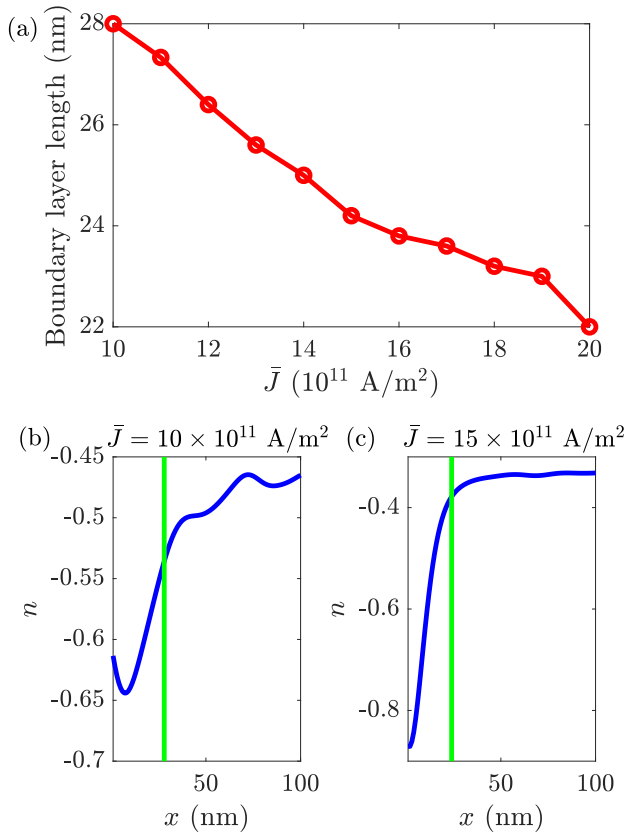}
\caption{ \label{fig:extra2} (color online) Boundary layer {width} as a function of the injection strength $\bar{u}$ }
\end{figure}

We observe a remarkable qualitative agreement between the micromagnetic simulations and the analytical results shown in Fig.~\ref{fig2}. In particular, we observe DEFs that follow the UHS isofrequency contours obtained in Sec. III without non-local dipole and in-plane anisotropy (solid blue curves) and CS-DEFs when the injection conditions are supersonic (solid red curves). Only three CS-DEFs are shown for clarity. The corresponding frequencies are shown in Fig.~\ref{fig5}(b) in physical units as a function of $\bar{J}$ and color-coded as {in} panel (a). We emphasize that a linear relation between $\bar{J}$ and $\bar{u}$ is not possible to obtain because of the particularities {of} the energy landscape imposed by the magnetization texture, anisotropy, and non-local dipole fields. Nonetheless, a maximum frequency is observed at the transition between DEFs and CS-DEFs.

The boundary layer {width} is difficult to calculate in micromagnetic simulations. This is because the frequency does not match exactly to the analytical results when nonlocal dipole and anisotropy fields are included and, therefore, the determination of the parameter $n_\infty$ is inaccurate. However, we can estimate the boundary layer size from the spatial profile of $n$. We determine the boundary layer {width} as the region in space where the slope of $n$ is larger than a threshold value of $0.005$ {in units of $1/\lambda_\mathrm{ex}$}. In Fig.~\ref{fig:extra2}(a), we show the boundary layer {width} as a function of injection current $\bar{J}$. The decreasing trend qualitatively agrees with the analytical results presented in Fig.~\ref{fig:extra} and the boundary layer {width} is within the predicted values for Py. Because the criterion used for {spin injection in} micromagnetic simulations is different than the analytical {boundary conditions}, we show the profile of $n$ close to the injection site for $\bar{J}=10\times10^{11}$~A/m$^2$ and $15\times10^{11}$~A/m$^2$ in Fig.~\ref{fig:extra2}(b) and (c), respectively. A reasonable estimation of the boundary layer {width}, c.f. with Fig.~\ref{fig1}, is observed.

\section{Conclusions}

In this paper, we analytically determined the form and qualitative features of magnetization states sustained by spin injection of arbitrary strength in ferromagnetic channels with easy-plane anisotropy. For this, we utilize a dispersive hydrodynamic formulation that captures the necessary physical terms without approximations while being analytically tractable. Our analytical study fully characterizes the possible solutions that support long-distance spin transport under a unified framework.

We find that DEFs are generally nonlinear in profile and frequency tuneability. Additionally, we characterize a novel solution, a CS-DEF, composed of a stationary soliton nucleated at the injection site that smoothly transitions into a nonlinear DEF. 
A notable consequence of the onset of CS-DEFs is that the frequency redshifts to injection. This feature is important for spintronic applications because it leads to a saturation of frequency and, therefore, of spin current magnitudes pumped {into} adjacent spin reservoirs. It is numerically found that the maximum frequency monotonically decays {with} the channel's length, indicating the increased energy that must be invested in the nonlocal compensation of damping to sustain DEFs. In other words, there is a compromise between the spin transport capacity and the length of the channel.

The adiabatic UHS interpretation introduced in this paper allows one to utilize the UHS phase {space's isofrequency contours} as a chart to categorize the magnetization states sustained in a ferromagnetic channel. This chart could be utilized to explore the profile of magnetization states induced in channels with two or more boundary conditions, e.g., contacts for STT and adjacent spin current reservoirs~\cite{Schneider2018}. The methodology presented here will be valuable for further analytical and numerical studies as well as to aid the design of an experimental realization of extended magnetization textures for microscopic spin transport.

\begin{acknowledgments}
The authors thank T. J. Silva for {helpful and insightful} discussions. This material is based on work supported by the U.S. Department of Energy, Office of Science, Office of Basic Energy Sciences under Award Number 0000231415. M.A.H. partially supported by NSF CAREER DMS-1255422.
\end{acknowledgments}

\appendix

\section{Asymptotic analysis}
\label{sec:asymptotic-analysis}

In this appendix, we implement an asymptotic analysis of the nonlinear
ordinary differential equations (ODEs) (\ref{eq:33}) and (\ref{eq:34})
subject to the boundary conditions (\ref{eq:61}) and (\ref{eq:62})
that leads to the dissipative exchange flow regimes identified in this
work: linear, nonlinear, and contact DEF solutions.

For this, we introduce the spatial rescaling
\begin{equation}
  \label{eq:6}
  y = \frac{x}{L},
\end{equation}
and use equation (\ref{eq:33}) to simplify equation (\ref{eq:34}) {and}
obtain the equivalent ODEs
\begin{subequations}
  \label{eq:14}
  \begin{eqnarray}
    \label{eq:16}
    0 &=& \left[(1-n^2)u\right]'+\alpha L (1-n^2)\tilde{\Omega},\\ 
    \label{eq:17}
    \tilde{\Omega} &=&
                       -(1-u^2)n+\frac{n''(1-n^2) +  n (n')^2}{L^2(1-n^2)^2},
  \end{eqnarray}
\end{subequations}
where the prime $'$ denotes a spatial derivative with respect to $y$,
$\tilde{\Omega} = \Omega/(1+\alpha^2)$ as before, and the boundary
conditions (\ref{eq:18}) become
\begin{subequations}
  \begin{eqnarray}
    \label{eq:15}
    n'(0)=0, &\quad& n'(1)=0,\\
    \label{eq:19}
	u(0)=\bar{u}, &\quad& u(1)=0 .
  \end{eqnarray}
\end{subequations}

\subsection{Linear DEF solution:  weak injection}
\label{sec:linear-def-solution}

The parameter regime that leads to the linear DEF solution requires
sufficiently weak injection, therefore we introduce the small
parameter $0 < |\bar{u}| \ll 1$ and the asymptotic expansions
\begin{equation}
  \label{eq:20}
  \begin{split}
    u &= \bar{u} u_1 + \bar{u}^3 u_3 + \cdots, \quad n = \bar{u} n_1 +
    \bar{u}^3 n_3 + \cdots, \\
    \tilde{\Omega} &= \bar{u} \tilde{\Omega}_1 + \bar{u}^3
    \tilde{\Omega}_3 + \cdots, \quad 0 < |\bar{u}| \ll 1.
  \end{split}
\end{equation}

Inserting them into equations (\ref{eq:14}), and equating like powers
of $\overline{u}$, we obtain the two equations
\begin{subequations}
  \label{eq:21}
  \begin{eqnarray}
    \label{eq:22}
    u_1' &=& - \alpha L \tilde{\Omega}_1, \\
    \label{eq:23}
    \frac{1}{L^2} n_1'' - n_1 &=& \tilde{\Omega}_1,
  \end{eqnarray}
\end{subequations}
at leading order ${\cal O}(\bar{u})$.  The boundary conditions
(\ref{eq:19}) and Eq.~(\ref{eq:22}) imply $u_1(y) = 1-y$,
$\tilde{\Omega}_1 = 1/(\alpha L)$.  The boundary conditions
(\ref{eq:15}) and Eq.~(\ref{eq:23}) imply $n_1(y) = -1/(\alpha L)$.
Inserting this approximate leading order solution into the expansions
(\ref{eq:20}) give the linear DEF solution (\ref{eq:def}).

We note that equating the next order terms ${\cal O}(\bar{u}^3)$ in
Eqs.~(\ref{eq:14}) leads to
\begin{subequations}
  \label{eq:A24}
  \begin{eqnarray}
    \label{eq:25}
    u_2' &=& \alpha L ( n_1^2\tilde{\Omega}_1-
                         \tilde{\Omega}_2)  + (n_1^2 u_1)' , \\
    \label{eq:A26}
    \frac{1}{L^2} n_2'' - n_2 &=& \tilde{\Omega}_2 -
                                  u_1^2 n_1 -
                                  \frac{1}{L^2} n_1
                                  (n_1')^2 .
  \end{eqnarray}
\end{subequations}
Inserting the leading order solution for $n_1$, $u_1$, and
$\tilde{\Omega}_1$ into Eq.~\eqref{eq:25} results in $u_2' = -\alpha L
\tilde{\Omega}_2$.  Applying the boundary conditions \eqref{eq:19}
($u_2(0) = u_2(1) = 0$) imply $u_2(y) = 0$ and $\tilde{\Omega}_2 = 0$.
Equation \eqref{eq:A26} and the boundary conditions \eqref{eq:15}
($n_2'(0) = n_2'(1) = 0$) are solved with a spatially varying $n_2(y)$
(superposition of exponentials and a quadratic polynomial in $y$).
This means that the linear DEF solution (\ref{eq:def}) approximates
the velocity and frequency to high accuracy, ${\cal O}(\bar{u}^5)$,
but the density has a spatially varying correction that scales with
$\bar{u}^3$.

It is important to note that the linear DEF solution only requires
sufficiently weak injection.  Inspection of the asymptotic solution
implies $0 < |\bar{u}| \ll \min(1,\alpha L)$ in order to maintain a
well-ordered asymptotic series in the
expansions~\eqref{eq:20}. Notably, there is no assumption on the
magnitude of the damping coefficient $\alpha$ nor channel length $L$.

\subsection{Nonlinear DEF solution:  long channel, subsonic injection}
\label{app:def}

In this section, we provide the detailed derivation of
Eqs.~\eqref{eq:8}, \eqref{eq:82}, and \eqref{eq:83}.  The assumption
of weak injection for the linear DEF solution is relaxed and now we
require a long channel, i.e., $L \gg 1$.  To this end, we assume the
asymptotic expansions
\begin{equation}
  \label{eq:24}
  \begin{split}
    u &= u_0 + \frac{1}{L^2} u_2 + \cdots, \quad n = n_0 +
    \frac{1}{L^2} n_2 + \cdots, \\  
    \tilde{\Omega} &= \tilde{\Omega}_0 + \frac{1}{L^2}
    \tilde{\Omega}_2 + \cdots, \quad L \gg 1,
  \end{split}
\end{equation}
insert them into equations \eqref{eq:14} and obtain the leading order
equations
\begin{eqnarray}
  \label{eqA:1}
  0 &=& u_0' - \frac{2u_0 n_0 n_0'}{1-n_0^2} + \alpha L\tilde{\Omega}_0, \\
  \label{eq:B26}
  \tilde{\Omega}_0 &=& -(1-u_0^2)n_0 .
\end{eqnarray}

Using Eq.~\eqref{eq:B26}, we can eliminate $n_0$ from Eq.~\eqref{eqA:1}
to obtain an ODE for $u_0$
\begin{equation}
  \label{eqA:3}
  \alpha L \tilde{\Omega}_0 =
  u_0' \left[
    \frac{4\tilde{\Omega}_0^2
      u_0^2}{(1-u_0^2)(u_0^4-2u_0^2+1-\tilde{\Omega}_0^2)} - 1\right],  
\end{equation}
which is equivalent to Eq.~\eqref{eq:73} in the main text. To
integrate this expression, we perform partial fraction
decomposition
\begin{eqnarray}
  \label{eqA:4}
  \alpha L \tilde{\Omega}_0 &=&  u'_0 \bigg[ -1+ \frac{4}{u_0^2-1}
  \\&& -\frac{2(1-\tilde{\Omega}_0)}{u_0^2-(1-\tilde{\Omega}_0)}
  - \frac{2(1+\tilde{\Omega}_0)}{u_0^2-(1+\tilde{\Omega}_0)}
  \bigg].\nonumber
\end{eqnarray}
This solution must agree with the linear DEF solution when $|\bar{u}|$
is small so, from Eq.~\eqref{eq:def}, we expect $|\tilde{\Omega}_0| <
1$ and we can integrate each term in Eq.~\eqref{eqA:4} to obtain an
implicit expression for $u_0(y)$
\begin{eqnarray}
  \label{eqA:6}
  \alpha L \tilde{\Omega}_0 y +C &=& -u_0 -4\tanh^{-1}{u_0} \\&+&
  2\sqrt{1-\tilde{\Omega}_0}
  \tanh^{-1}{\left(\frac{u_0}{\sqrt{1-\tilde{\Omega}_0}} \right)}
  \nonumber \\&+&2 \sqrt{1+\tilde{\Omega}_0}
  \tanh^{-1}{\left(\frac{u_0}{\sqrt{1+\tilde{\Omega}_0}}
    \right)}\nonumber ,
\end{eqnarray}
where $C$ is an integration constant.  Evaluating the boundary
condition \eqref{eq:19} ($u_0(1)=0$), we obtain the integration
constant
\begin{equation}
  \label{eqA:7}
  C = - \alpha\tilde{\Omega}_0L .
\end{equation}
Replacing $C$ in Eq.~\eqref{eqA:6}, we obtain the implicit solution
for the fluid velocity that is given in the main text as
Eq.~\eqref{eq:8}.  The frequency $\Omega_0$ in Eq.~\eqref{eq:82} and
density $n_0$ in Eq.~\eqref{eq:83} follow from the boundary condition
$u_0(0) = \bar{u}$ and Eq.~\eqref{eq:B26}, respectively.

This implicit solution satisfies the boundary conditions for the
velocity \eqref{eq:19} but it only satisfies $n_0'(1) = 0$ and not
$n_0'(0) = 0$.  While this could be resolved by considering a boundary
layer adjacent to $y = 0$, the fact that we are considering a long
channel implies $\frac{d}{dx} n_{\rm DEF}(0) = \mathcal{O}(L^{-1})$,
which is negligibly small within the asymptotic approximation
\eqref{eq:24}.

It is worth noting that the asymptotic expansion of the nonlinear DEF
solution for small $|\overline{u}|$ is
\begin{eqnarray}
  \label{eq:27}
  u_{\rm DEF}(x) &=& \bar{u}\left ( 1- \frac{x}{L} \right ) \\
  &+& \bar{u}^5 \frac{4 \frac{x}{L}
    \left ( 1 - \frac{x}{L} \right )}{3 (\alpha L)^2} \left (
    \frac{x}{L} - 2 \right ) + \mathcal{O}(\bar{u}^7) , \nonumber \\
  \label{eq:28}
  n_{\rm DEF}(x) &=& - \frac{\bar{u}}{\alpha L} - \bar{u}^3
  \frac{\left ( 1 - \frac{x}{L} \right )^2}{\alpha L } +
  \mathcal{O}(\bar{u}^5) ,\\
  \label{eq:29}
  \tilde{\Omega}_{\rm DEF} &=& \frac{\bar{u}}{\alpha L} - \frac{4
    \bar{u}^5}{3(\alpha L)^3} + \mathcal{O}(\bar{u}^7) ,
\end{eqnarray}
which agrees with the linear DEF solution at leading order and at
$\mathcal{O}(\bar{u}^3)$ for $L \gg 1$.  A useful result is obtained
by evaluating the nonlinear DEF solution at $x = 0$, which gives the
relationship
$n_\mathrm{DEF}(0) = -(\bar{u} + \bar{u}^3)/(\alpha L) +
\mathcal{O}(\bar{u}^5)$ between the spin density at the injection site
and the injection velocity.

Although we have assumed $L \gg 1$, we have made no assumption on
magnetic damping $\alpha$.  As noted in Sec.~\ref{sec:def-cs-def}, the
DEF frequency $\tilde{\Omega}$ saturates when injection achieves the
local speed of sound [Eq.~\eqref{eq:9}].  This sets the maximum
injection $\bar{u}$---which can still be relatively large---for the
nonlinear DEF solution, i.e., injection must be \textit{subsonic}.

\subsection{CS-DEF solution:  weak damping, long channel, supersonic injection}
\label{app:bl}

In order to investigate the supersonic injection regime, we need to
introduce a boundary layer near $y = 0$ in Eqs.~\eqref{eq:14} (see,
e.g., Ref.~\onlinecite{Bender1999}).  For this, we consider two
separate solution regions: an inner region close to the injection site
and an outer region that extends to the unforced edge of the channel.
The solutions from these two regions are then asymptotically matched
in order to obtain a uniformly valid asymptotic approximation across
the entire channel.

\subsubsection{Inner region}

In the inner region, we are interested in the solution profile close
to $y = 0$.  Therefore, we ``zoom'' into this region for
Eqs.~\eqref{eq:14} by returning to the $x = y L$ scale \eqref{eq:6}
where $L$ is assumed large
\begin{subequations}
  \label{eqB:1}
  \begin{eqnarray}
    \label{eqB:11v1}
    0 &=& \left[(1-n^2)u\right]'+\alpha (1-n^2)\tilde{\Omega},\\ 
    \label{eqB:12v1}
    \tilde{\Omega} &=&
                       -(1-u^2)n+\frac{n''(1-n^2) +  n
                         (n')^2}{(1-n^2)^2} .
  \end{eqnarray}
\end{subequations}
Now, the prime $'$ is a derivative with respect to $x$.  As we will
see, only the leftmost boundary conditions in \eqref{eq:15} and
\eqref{eq:19} will be satisfied in the inner region.  Anticipating the
behavior of the solution in the outer region that we will match to, we
use the following boundary conditions
\begin{subequations}
  \label{eq:30}
  \begin{eqnarray}
    \label{eqB:41}
    n'(0)=0, &\quad& \lim_{x \to \infty} n'(x)= n_\infty , \\
    \label{eqB:42}
	u(0)=\bar{u}, &\quad& \lim_{x \to \infty} u(x)= u_\infty,
  \end{eqnarray}
\end{subequations}
with $n_\infty$ and $u_\infty$ to be determined.

To approximately solve Eqs.~\eqref{eqB:1} subject to the boundary
conditions \eqref{eq:30}, we assume weak damping $0 < \alpha \ll 1$
and expand in the asymptotic series
\begin{equation}
  \label{eq:BB24}
  \begin{split}
    u &= u_0 + \alpha u_1 + \cdots, \quad n = n_0 +
    \alpha n_1 + \cdots, \\  
    \tilde{\Omega} &= \tilde{\Omega}_0 + \alpha
    \tilde{\Omega}_1 + \cdots, \quad 0 < \alpha \ll 1 .
  \end{split}
\end{equation}
This implies that in the inner region, the dynamics are effectively
conservative to leading order
\begin{subequations}
\label{eqB:3}
\begin{eqnarray}
  \label{eqB:31}
  0 &=& \left[(1-n_0^2)u_0\right]',\\ 
  \label{eqB:32}
    \tilde{\Omega} &=&
    -(1-u_0^2)n_0+\frac{n_0''(1-n_0^2) +  n_0
      (n_0')^2}{(1-n_0^2)^2} .
  \end{eqnarray}
\end{subequations}

To continue, we integrate Eq.~\eqref{eqB:31} to obtain $u_0$ in terms
of $n_0$
\begin{equation}
  \label{eq:35}
  u_0 = \frac{C}{1 - n_0^2},
\end{equation}
where $C$ is a constant of integration.  We substitute this into
Eq.~\eqref{eqB:32} and multiply by $2n'$ to obtain
\begin{equation}
  \label{eqB:5}
  2 \tilde{\Omega}_0 n_0'+ 2n_0n_0' - C^2 \left ( \frac{1}{1-n_0^2} \right )' =
  \left[\frac{1}{1-n_0^2}\left(n_0'\right)^2\right]' .
\end{equation}
Every term in Eq.~\eqref{eqB:5} is a perfect derivative. Therefore,
upon integration, we obtain the first order ODE
\begin{equation}
  \label{eqB:6}
  \left( n_0' \right)^2 = -n_0^4 - 2\tilde{\Omega}_0 n_0^3+ (1-K)n_0^2+
  2\tilde{\Omega}_0n_0- C^2+K,
\end{equation}
where $K$ is an additional constant of integration.  This ODE can
generally be integrated in terms of elliptic integrals (see, e.g.,
Ref.~\onlinecite{Congy2016}) but we are interested in the localized,
stationary soliton solution that satisfies the boundary conditions
\eqref{eq:30}, which is
\begin{subequations}
  \label{eqB:7}
  \begin{eqnarray}
    \label{eqB:71}
    n_{in} &=& \frac{a\nu_1\mathrm{tanh}^2{(\theta
               x)}+\nu_2(n_\infty-a)}{a\mathrm{tanh}^2{(\theta x)}+\nu_2},\\ 
    \label{eqB:72}
    u_{in} &=& u_\infty\frac{1-n_\infty^2}{1-n_{in}^2},\\
    \label{eqB:73}
    \tilde{\Omega}_{in} &=& -n_\infty(1-u_\infty^2),
  \end{eqnarray}
\end{subequations}
where $\nu_1=a-n_\infty-2n_\infty u_\infty^2$,
$\nu_2=a-2n_\infty-2n_\infty u_\infty^2$,
$\theta=\sqrt{1-u_\infty^2-n_\infty^2(1+3u_\infty^2)}$, and
$a=n_\infty(1+u_\infty^2)+\sqrt{(1-u_\infty^2)(1-n_\infty^2
  u_\infty^2)}$.  The soliton's density deviation from its far-field
value $n_\infty$ is the amplitude $a$.  {Note that the soliton's extremum is situated at $x=0$ to enforce the BC $n'(0)=0$}. An additional relation is due
to spin injection at the left boundary $x = 0$ where the soliton's
extremum is attained
\begin{equation}
  \label{eqB:8}
  \bar{u} = u_{in}(0) = u_\infty\frac{1-n_\infty^2}{1-\left(n_\infty-a\right)^2}.
\end{equation}
This relation constrains $n_\infty$ and $u_\infty$.  We require an
additional relation to fully determine the solution.  This comes from
the asymptotic solution in the outer region, far from the forced
injection boundary at $x = 0$.

The soliton established in the inner region is therefore given by
Eqs.~\eqref{eqB:71}, \eqref{eqB:72}, and \eqref{eqB:8}, reported in
the main text.

\subsubsection{Outer region}

For the outer region, we return to the scaled variable $y = x/L$
\eqref{eq:6} and Eqs.~\eqref{eq:14}.  In order to match the inner
solution \eqref{eqB:7}, we need to modify the boundary conditions
\eqref{eq:17} to
\begin{subequations}
  \label{eq:36}
  \begin{eqnarray}
    \label{eqB:101}
    \lim_{y\rightarrow0} n(y)=n_\infty, &\quad& n'(1)=0,\\
    \label{eqB:102}
    \lim_{y\rightarrow0} u(y)=u_\infty, &\quad& u(1)=0 .
  \end{eqnarray}
\end{subequations}
The approximate outer solution to Eqs.~\eqref{eq:14} subject to the
boundary conditions \eqref{eq:36} is the nonlinear DEF solution
described in Sec.~\ref{app:def} with $L \gg 1$, $\bar{u} \to
u_\infty$, which satisfies the following [cf.~Eqs.\eqref{eq:8},
\eqref{eq:82}, \eqref{eq:83}]
\begin{subequations}
  \label{eq:38}
  \begin{eqnarray}
    \label{eq:39}
    \alpha L \tilde{\Omega}_{out} \left(1-y \right)
    &=& u_{out} +4\tanh^{-1}{(u_{out})}  \\&-&2
    \left[\mathcal{N}^-(u_{out},\tilde{\Omega}_{out})
      +\mathcal{N}^+(u_{out},\tilde{\Omega}_{out})\right], \nonumber \\
    \label{eq:40}
    n_{out}(y) &=& - \frac{\tilde{\Omega}_{out}}{1 - u_{out}(y)^2}, \\
    \label{eq:41}
    \alpha L\tilde{\Omega}_{out} &=& u_\infty + 4\tanh^{-1}{(u_\infty)}
     \\&-&2
    \left[\mathcal{N}^-(u_\infty,\tilde{\Omega}_{out}) +
      \mathcal{N}^+(u_\infty,\tilde{\Omega}_{out})\right],  \nonumber
  \end{eqnarray}
\end{subequations}
However, the boundary condition $n'(0) = 0$ no longer applies.  Instead, we have a fixed value of the
spin density 
\begin{equation}
  \label{eq:42}
  n_\infty = n_{out}(0) = - \frac{\tilde{\Omega}_{out}}{1 -
    u_\infty^2} .
\end{equation}
This relation and Eq.~\eqref{eqB:73} imply the equality of the inner
and outer precessional frequencies so we define
\begin{equation}
  \label{eq:37}
  \tilde{\Omega} = \tilde{\Omega}_{in} = \tilde{\Omega}_{out} .
\end{equation}

\subsubsection{Matching}

The full solution for the CS-DEF is obtained by matching the inner
solution to the outer solution.  Actually, the choice of boundary
conditions in Eqs.~\eqref{eq:30} and \eqref{eq:36} encodes the
matching of the two solutions.  We now summarize the three equations
that uniquely determine $n_\infty$, $u_\infty$, and $\tilde{\Omega}$
in terms of the spin injection $\overline{u}$.  They are
\begin{subequations}
  \label{eq:43}
  \begin{eqnarray}
    \label{eq:44}
    \tilde{\Omega} &=& -n_\infty(1 - u_\infty^2),  \\
    \label{eq:45}
    \bar{u} &=& \frac{u_\infty(1-n_\infty^2)}{1 - (n_\infty - a)^2},  \\
    \label{eq:46}
    \alpha L \tilde{\Omega} &=& u_\infty + 4 \tanh^{-1}{(u_\infty)}
     \\ &-& 2
     \left[\mathcal{N}^-(u_\infty,\tilde{\Omega}) +
       \mathcal{N}^+(u_\infty,\tilde{\Omega})\right] \nonumber,
  \end{eqnarray}
\end{subequations}
coinciding with Eqs.~\eqref{eqB:73}, \eqref{eqB:8}, and \eqref{eq:41},
respectively.

With all parameters determined, we can now obtain a uniformly valid
asymptotic approximation to the CS-DEF with
\begin{subequations}
  \label{eqB:14}
  \begin{eqnarray}
    \label{eqB:141}
    u_{cs}(x) &=& u_{in}(x) + u_{out}(x/L) - u_\infty,\\
	\label{eqB:142}
    n_{cs}(x) &=& n_{in}(x) + n_{out}(x/L) - n_\infty,
  \end{eqnarray}
\end{subequations}
which is the approximation used, for example, in Fig.~\ref{fig1}(b).

\end{document}